\documentclass[useAMS,usenatbib]{mn2e} 
\usepackage{graphicx}

\begin{document}

\title[Maximum-Likelihood combination of Cross-Correlations]
{Cross-Correlation and Maximum Likelihood Analysis: \\
A New Approach to Combine Cross-Correlation Functions}

\author[S. Zucker]{S. Zucker \\
Dept. of Geophysics and Planetary Sciences and Wise
  Observatory, Raymond and Beverly Sackler Faculty of Exact Sciences, \\
  Tel Aviv University, Tel Aviv, Israel}

\maketitle

\begin{abstract} 
This paper presents a new approach to combine cross-correlation
functions.  The combination is based on a maximum-likelihood approach
and uses a non-linear combination scheme.  It can be effective for
radial-velocity analysis of multi-order spectra, or for analysis of
multiple exposures of the same object. Simulations are presented to
show the potential of the suggested combination scheme. The technique
has already been used to detect a very faint companion of HD\,41004.
\end{abstract}
  

\section{Introduction}

Ever since the seminal works of \citet{Sim1974} and
\citet{TonDav1979}, the cross-correlation technique to measure
astronomical Doppler shifts has become extremely popular.  The advent
of digitized spectra and computers made it the preferred method.
It has been applied in all the astronomical fields that require the
measurement of radial velocities from observed spectra, ranging from
binary and multiple
stellar systems to cosmology. In recent years, improvements in the
precision of radial velocities measured through cross-correlation led
to the detection of many extrasolar planets
\citep[e.g.,][]{MayQue1995}.

The cross-correlation technique is conceptually simple and can be
presented in an intuitive manner. Nevertheless, its properties 
 have been studied extensively, in an effort
to improve its precision and overcome its few limitations.  Thus,
various methods have been suggested to estimate its precision
\citep[e.g.,][]{TonDav1979,Con1985,MurHea1991}.
{\small TODCOR} - a Two-Dimensional Correlation technique, was
introduced as a generalization of cross-correlation meant to measure
the Doppler shifts of two blended spectra \citep{ZucMaz1994}.

Due to the progress in detector technology, the modern spectrographs
pose a new challenge to the technique, by producing multi-order
spectra. In order to maximize the precision of the measured
velocities, we need to combine the spectral information in all the
relevant orders, potentially reducing
 the signal-to-noise ($S/N$) required to achieve a
specified precision. One approach was suggested by \citet{Con1985}
and \citet{Bouetal2001}, who calculated the cross-correlation for each
order separately, and then calculated a weighted average of the
resulting cross-correlation functions. The weights used by
\citeauthor{Con1985} and \citeauthor{Bouetal2001} reflect their
estimate of the $S/N$ in the corresponding orders.

This paper offers another approach to study the properties of
cross-correlation. It is shown that under certain assumptions,
measuring the radial velocity through cross-correlation is equivalent
to using maximum-likelihood analysis for the measurement.  This
approach has been already used in communication theory in the context
of signal detection \citep[e.g.,][]{Pro1995}.  This approach leads in
a natural way to an error estimate and to a new scheme for combining
cross-correlation functions. This new scheme is shown, theoretically
and through simulations, to be superior over the other schemes
proposed, both in terms of precision and in terms of signal detection
capability.

Section \ref{maxlike} shows how cross-correlation can be derived from
maximum-likelihood theory. It derives an error-estimate and shows
simulations to test this error estimate.  The new combination scheme
is derived from maximum-likelihood principles in Section
\ref{combine}. Several simulations to demonstrate the power of the
technique are also presented in this Section.  The paper is concluded
by a few remarks in Section \ref{concl}.

\section{From maximum likelihood to cross-correlation}
\label{maxlike}

\subsection{The basic assumptions}

Let $f(n)$ denote the observed spectrum, whose Doppler shift is to be
found by correlating it against $g(n)$ -- the `template' of zero
shift. Both the stellar spectrum and the template are assumed to be
described as functions of the bin number -- $n$, where $n = A\ln
\lambda + B \ .$ Thus, the Doppler shift results in a uniform linear
shift of the spectrum \citep{TonDav1979}.
                   
Now let us introduce the statistical model.  Under this model we
assume that the observed spectrum was produced by multiplying the
template by a scaling constant ($a_0$), shifting it ($s_0$ bins) and
adding a random white gaussian noise with a fixed standard deviation
($\sigma_0$), i.e.:
\begin{eqnarray*}
f(n) &= &a_0 g(n-s_0) + d_n \ , \\ 
d_n &\sim &N(0,\sigma_0^2)
\end{eqnarray*}
Obviously, $a_0$ and $s_0$ are not known in advance. In the
literature, $\sigma_0$ is usually assumed to be known, but in the
following derivation we will assume no prior knowledge of $\sigma_0$.

A few simplifying assumptions are usually applied, which are also used
in this work.  Thus, the spectra involved are assumed to be
``continuum-subtracted'', i.e. a best-fit low-order polynomial has
been subtracted from the spectra. As a result, the spectra will have
a zero mean:
\begin{eqnarray*}
\sum_n f(n) &= &0 \\
\sum_n g(n) &= &0
\end{eqnarray*}
The number of bins is typically very large (usually in the order of
thousands) and thus, for small shifts (even tens of bins), one can
neglect edge effects and assume that for each shift, the overlap
length is $N$ -- the total number of bins.

\subsection{Maximum-likelihood estimation}

The likelihood is defined by the probability of the observed results,
under the assumed model, as a function of the model parameters. In our
case:
\begin{eqnarray*}
L &= &\prod_n \left ( \frac{1}{\sqrt{2 \upi \sigma^2}} \right )
\exp{\left \{ \frac{-[f(n)-a g(n-s)]^2}{2 \sigma^2} \right \} } \\
&= &\left ( \frac{1}{\sqrt{2 \pi \sigma^2}} \right ) ^N 
\exp{\left \{ -\sum_n \frac{[f(n)-a g(n-s)]^2}{2 \sigma^2} \right \} }
\end{eqnarray*}

The usual practice in maximum-likelihood estimation theory is to find
the values of the  parameters which maximize the natural
logarithm of the likelihood fuction. In our case this logarithm is: 
\[
\log{L} = -N \log{\sigma} - \frac{1}{2 \sigma^2} \sum_n [f(n)-a g(n-s)]^2 + 
\mathit{const.} \ ,
\] 
where $\mathit{const.}$ is independent of the parameters.  As it
turns out, the values of $\sigma$, $a$ and $s$ which maximize this
function are the solutions of the equations:
\begin{eqnarray*}
\hat{\sigma}^2 &= &\frac{1}{N} \sum_n [f(n) - \hat{a} g(n-\hat{s})]^2 \\
\hat{a} &= &\frac{\sum_n f(n) g(n-\hat{s})}{\sum_n g^2(n)}
\end{eqnarray*}
and $\hat{s}$ is the value which maximizes the cross-covariance function,
which is defined as:
\[
R(s) = \frac{1}{N} \sum_n f(n) g(n-s) \ .
\]

Let us introduce some notation which will facilitate the derivation
from here on. Thus, let $s_f$ and $s_g$ denote the standard deviation
of the analysed spectrum and the template:
\begin{eqnarray*}
s_f^2 &= &\frac{1}{N} \sum_n f^2(n) \\
s_g^2 &= &\frac{1}{N} \sum_n g^2(n) \ .
\end{eqnarray*}
Define $C(s)$ as the cross-correlation function of $f$ and $g$:
\[
C(s) = \frac{R(s)}{s_f s_g} \ ,
\]
where the normalization forces $|C(s)|$ to be smaller then unity. Since $C(s)$ is
proportional to $R(s)$, $\hat{s}$ can be defined as the shift which maximizes the
cross-correlation function.

Using these notations, the expressions for $\hat{\sigma}$ and
$\hat{a}$ reduce to:
\begin{eqnarray*}
\hat{\sigma}^2 &= &s_f^2 [1 - C^2(\hat{s})] \\
\hat{a} &= &\frac{R(\hat{s})}{s_g^2} \ .
\end{eqnarray*} 
Substituting these expressions again into the expression for $\log L$, we get:
\begin{equation}
\label{onelog}
\log L = -\frac{N}{2} \log{[1-C^2(\hat{s})]} + \mathit{const.}
\end{equation}

Thus, we see that the likelihood is an increasing monotonic function
of the squared cross-correlation. The dependence on the
square of the cross-correlation, instead of the cross-correlation
itself, means that the likelihood rises for negative values of the
correlation. This behaviour 
is related to the formal possibility that the scaling factor -- $a$ -- be
negative. This has no meaning in the astrophysical context, since
it means all the expected asborption lines in one component appear as
emission lines.

\subsection{Error estimate}

Maximum-likelihood theory also includes a way to estimate the errors,
or the confidence intervals, for the parameters. The covariance matrix
of the parameters has been shown, at non-pathological cases, to be the
negative inverse of the Hessian matrix of the log-likelihood function
at its maximum \citep[e.g.,][]{KenStu1967}. Let us calculate this
matrix and substitute the values at the maximum, treating the shift,
$s$, as a continuous variable:

\begin{eqnarray*}
\left. \frac{\upartial^2 \log{L}}{\upartial a^2} \right \vert _{\mathit maximum}
&= &-N \frac{s_g^2}{\hat{\sigma}^2} \\
\left. \frac{\upartial^2 \log{L}}{\upartial \sigma^2} \right \vert _{\mathit maximum}
&= &\frac{N}{\hat{\sigma}^2} - \frac{3}{\hat{\sigma}^4} 
   \sum_n[f(n) - \hat{a} g(n-\hat{s})]^2 = -\frac{2N}{\hat{\sigma}^2} \\
\left. \frac{\upartial^2 \log{L}}{\upartial s^2} \right \vert _{\mathit maximum}
&= & \frac{\hat{a} N R''(\hat{s})}{\hat{\sigma}^2} \\
\left. \frac{\upartial^2 \log{L}}{\upartial a \upartial \sigma} \right \vert _{\mathit maximum}
&= & \frac{2N}{\hat{\sigma}^3}[R(\hat{s}) - \hat{a} s_g^2] = 0 \\
\left. \frac{\upartial^2 \log{L}}{\upartial a \upartial s} \right \vert _{\mathit maximum}
&= & -\frac{N R'(\hat{s})}{\hat{\sigma}^2} = 0 \\
\left. \frac{\upartial^2 \log{L}}{\upartial \sigma \upartial s} \right \vert _{\mathit maximum}
&= & -\frac{2 \hat{a} N R'(\hat{s})}{\hat{\sigma}^2} = 0 \ ,
\end{eqnarray*}
where $R'(s)$ and $R''(s)$ are the first and second derivatives of the
cross-covariance function.  The last two equations follow from the
definition of $\hat{s}$ as the shift where the maximum correlation is
achieved.

Thus, we see that the Hessian matrix is, fortunately, diagonal and the
inversion can be accomplished easily. The resulting squared error is:
\[
\sigma_s^2 = -\frac{\hat{\sigma}^2}{\hat{a} N R''(\hat{s})} =
-\bigl[N \frac{C''(\hat{s})}{C(\hat{s})} \frac{C^2(\hat{s})}{1-C^2(\hat{s})}\bigr]^{-1} \ ,
\]
where $C''(s)$ is the second derivative of the cross-correlation
function.

It is instructive to examine the above expression. It is factored into
three separate factors: $N$, the number of bins, which has an
inverse relation to the error, as is expected intuitively;
$\frac{C''(\hat{s})}{C(\hat{s})}$, which is a normalized
measure of the sharpness of the cross-correlation peak, and it depends
very weakly on the $S/N$.  A larger absolute value of
$\frac{C''(\hat{s})}{C(\hat{s})}$ implies a sharper peak and
therefore a smaller error, and vice versa. The third factor,
$\frac{C^2(\hat{s})}{1-C^2(\hat{s})}$ is a measure of the ``line''
$S/N$ ratio of the spectrum, i.e., the ratio between the signal
standard-deviation and that of the noise. This last statement can be
easily seen by the relation of $1-C^2(\hat{s})$ to the noise standard
deviation.

A different error estimate would result if the template is also taken
to be a noisy spectrum, as is sometimes assumed in the literature
\citep[e.g.,][]{VerDav1999} and thus the procedure would have to
estimate the 'true' template using the spectrum and the template. The
formal calculations are much more complicated in this case and the
resulting error estimate is approximately $\sqrt{2}$ times larger than
the one quoted above. However, in the usual practice, where the same
template is used for many observed spectra, the model presented here
is more suitable, since the scatter of the estimated velocities would
be influenced only by the noise in the spectra, whereas the noise in
the template would contribute a systematic error.

An important feature of maximum-likelihood estimation is that
asymptotically (for large $N$), it produces a ``minimum-variance
estimator'' \citep{KenStu1967}. This means we cannot expect to achieve
an estimate with smaller errors. Since we have shown that the
cross-correlation estimate of the shift is equivalent to a
maximum-likelihood estimate, this shows the optimality of the
cross-correlation method. It is important to bear in mind, though,
that the statistical model we used was very simplistic, and other
techniques may prove better for removing effects not included in this
model, such as spectral mismatch or instrumental long-term trends
\citep[e.g.,][]{Che2000}.

\subsection{Simulations}
\label{simulations_1}

\begin{figure}
\resizebox{\hsize}{!}{\includegraphics{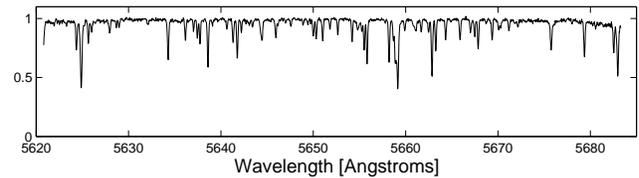}}
\caption{ A single order from the observed multi-order spectrum of
HD\,38858.  This order is the one used as a template for the error
analysis simulation.}
\label{template}
\end{figure}

The simulations presented here aim to test the above error estimate
under controlled conditions. All the simulations use a spectrum of the
G-dwarf HD\,38858, obtained by {\small CORALIE}, as part of a program
to derive precise abundances of planet-hosting and non-planet-hosting
stars \citep{Sanetal2000,Sanetal2001}.  To test the error estimate a
single spectral order was used, ranging from $5\,621$\ to
$5\,683$\,\AA, with $1\,920$ bins.  Fig.~\ref{template} presents this
template spectrum.  In order to simulate the above specified
statistical model, care was taken to eliminate other effects which are
not tested in this work. Thus, the assumed doppler shift was always
$0$, to avoid edge-effects, and the spectrum was rectified to a
constant continuum level of unity.

\begin{figure}
\resizebox{\hsize}{!}{\includegraphics{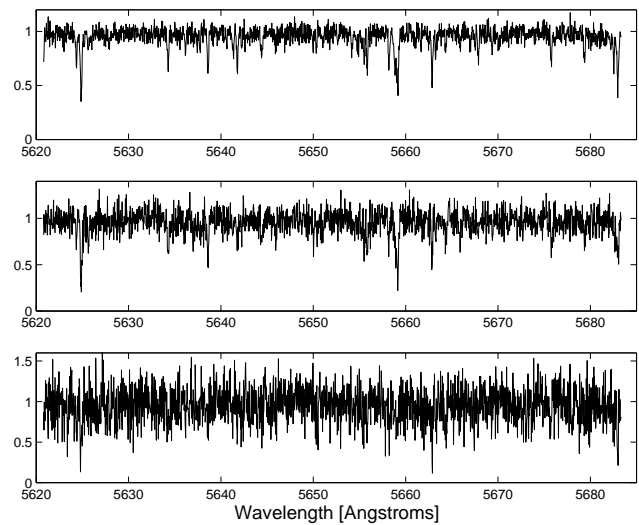}}
\caption{ 
Three simulated spectra, using the template shown in
Fig.~\ref{template}.  The noise levels used for the simulation are
$0.05$ (top), $0.1$ (middle) and $0.2$ (bottom).}
\label{spectra}
\end{figure}

\begin{figure}
\resizebox{\hsize}{!}{\includegraphics{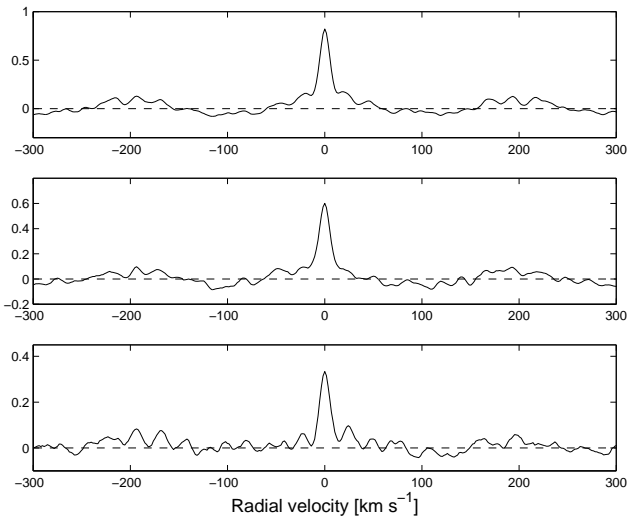}}
\caption{ 
The cross-correlation functions between the spectra shown in
Fig.~\ref{spectra} and the template (Fig.~\ref{template}).}
\label{xcorrs}
\end{figure}

\begin{figure}
\resizebox{\hsize}{!}{\includegraphics{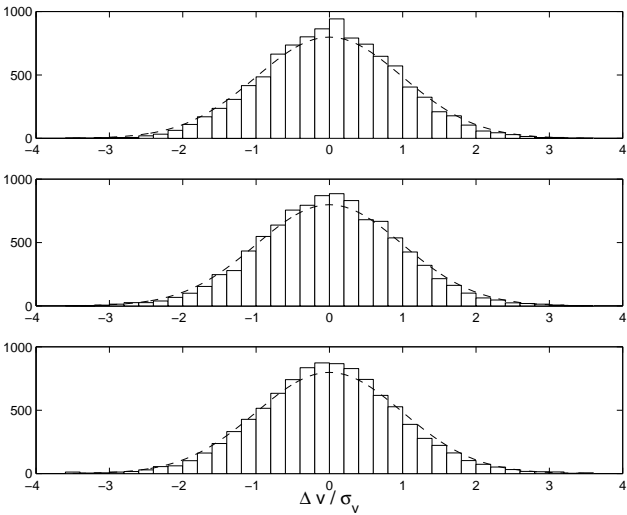}}
\caption{ 
Histograms of the ratio between the offset of the cross-correlation
function ($\Delta v$) and the estimated error ($\sigma_v$). The three
panels correspond to the three noise levels used in Fig.~\ref{spectra}
and \ref{xcorrs}. The dashed lines represent a zero-mean unit-variance
Gaussian distribution.}
\label{histograms}
\end{figure}

The noise added to the spectra was a simple Gaussian white noise, with
standard deviations of $0.05$, $0.1$ or $0.2$.  Sample noisy spectra
are shown in Fig.~\ref{spectra}. These spectra were cross-correlated
against the known template, resulting in the cross-correlation
functions presented in Fig.~\ref{xcorrs}.  The exact peak locations
and the relevant derivatives were estimated by interpolating with a
parabola around the peak.  Each simulation was repeated $10\,000$
times. Figure \ref{histograms} shows the distribution of the ratio
between the actual offset of the cross-correlation peak ($\Delta v$)
and the error estimate ($\sigma_v$).  The figure shows that this ratio
is distributed as a zero-mean unit-variance Gaussian for all the
noise-levels tested, as we would expect for a valid error estimate.
The apparent Gaussian nature of the distribution, and especially its
unimodality, show the regularity of the measurement error and its
estimate. In later sections we will use the same procedure to test the
error estimate of the proposed combination scheme and expect to obtain
similar distributions.

\section{Combining cross-correlation functions}
\label{combine}

\subsection{The combination scheme}

Suppose we now have $M$ separate spectra, each one corresponding to a
separate template, and all are assumed to have the same Doppler
shift. These separate spectra can either be separate exposures of the
same object, or separate orders of the same exposure. The following
analysis assumes they all possess the same number of bins -- $N$,
although this assumption can be very easily relaxed. We do not assume
the same $S/N$ ratios for the various spectra nor the same scaling
constants.

The most simple scheme to combine cross-correlation functions into one
single function is the straightforward, non-weighted average
($\rmn{SA}$):
\[
\rmn{SA}(s) = \frac{1}{M} \sum_{i=1}^{M} C_i(s) \ ,
\]
where $M$ is the number of cross-correlation functions to be combined,
while $C_i(s)$ is the $i$-th cross-correlation, calculated between
the $i$-th spectrum and the $i$-th template.

Another simple combination scheme is based on the so-called
``determination coefficient'', i.e., the squared correlation
coefficient -- $C^2(s)$ in our case. This quantity is commonly used in
regression analysis to describe the degree to which the variability in
one variable is explained by the other \citep[e.g.,][]{Hay1988}, and
is supposed to be additive.  Therefore, we may try to combine these
determination coefficients ($\rmn{DC}$) to get:
\[
\rmn{DC}^2(s) = \frac{1}{M} \sum_i C_i^2(s) \ .  
\]
This value can be interpreted as a weighted sum of the correlations.
Each term is weighted by itself, since its value - the correlation -
is also a measure of its reliability.

The likelihood approach can provide another scheme to combine
cross-correlations. First, let us see what is the overall likelihood
of the observations:
\[
L = \prod_i \left ( \frac{1}{\sqrt{2 \upi \sigma_i^2}}
\right ) ^N
\exp{\left \{- \frac{1}{2 \sigma_i^2} \sum_n \left [f_i(n) - a_i g_i(n-s) \right ]^2 \right \} }
\]
Converting, as usual to the logarithm, we get:
\[
\log L = - \sum_i \left \{ N \log \sigma_i - 
\frac{1}{2 \sigma_i^2} \sum_n \left [ f(n) - a_i g_i(n-s) \right ]^2 \right \}
+ \mathit{const.}
\]
Substituting the optimal values of $\sigma_i$ and $a_i$ leads eventually to:
\[
\log L = - \frac{N}{2} \sum_i \log \left [1 - C_i^2(s)\right ] + \mathit{const.}
\]
which is very similar to eq. \ref{onelog}. Therefore we can equate these
two expressions to get an ``effective'' correlation value -- $\rmn{ML}$:
\[
N M \log \left [1 - \rmn{ML}^2(s)\right ] = \sum_i N \log \left [ 1 - C_i^2(s)\right ]
\]
or:
\[
\rmn{ML}^2(s) = 1 - \left \{\prod_i \left [1-C_i^2(s)\right ] \right \}^{1/M}
\]

As it turns out, for sufficiently small $C_i(s)$ the above expression
reduces to the expression of the ``determination coefficient'' mean --
$\rmn{DC}(s)$. On the other extreme, if one $C_i(s)$ approaches
unity (meaning it has a very high $S/N$), $\rmn{DC}(s)$ will
approach unity.  This can be understood intuitively, since the other
cross-correlations probably represent much poorer $S/N$ and therefore
can be neglected.

In order to obtain an error estimate for the shift, the Hessian of the likelihood has
to be calculated and inverted. This time, the Hessian is not diagonal, but a 
laborious calculation yields the following expression for the error:
\[
\sigma_s^2 = -\bigl[M N \frac{\rmn{ML}''(\hat{s})}{\rmn{ML}(\hat{s})} 
   \frac{\rmn{ML}^2(\hat{s})}{1-\rmn{ML}^2(\hat{s})} \bigr]^{-1} \ ,
\]
which is remarkably similar to the expression for the error in the single spectrum case, except for
using the total number of bins -- $M N$ instead of $N$! 

\subsection{Simulations}

\begin{figure}
\resizebox{\hsize}{!}{\includegraphics{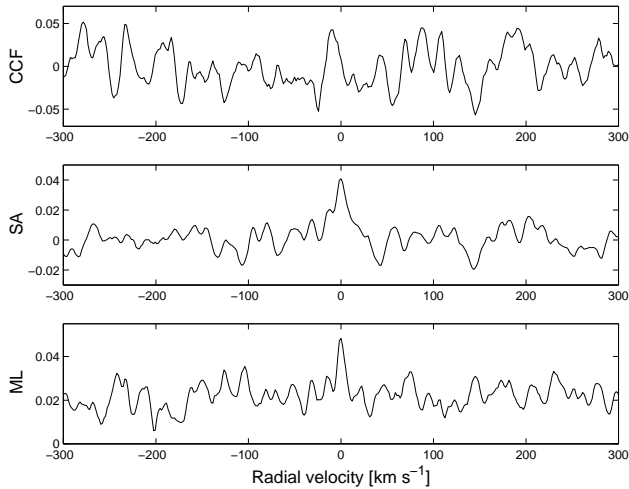}}
\caption{ 
Top: Cross-correlation function corresponding to a single order, with
a noise level of $3.5$.  Middle: A simple average of the
cross-correlations of $10$ orders, all with a noise level of $3.5$.
Bottom: maximum-likelihood combination of the same $10$
cross-correlations.}
\label{demo1}
\end{figure}

In order to demonstrate the effect of combining cross-correlations,
$10$ spectral orders were used, taken from the same spectrum used in
\ref{simulations_1}, covering the spectral range from $5\,189$\ to 
$5\,683$\,\AA.  As a first test, noise was added to these $10$ orders,
with a very high standard deviation of $3.5$. Figure \ref{demo1}
presents the results of this test.  The top panel shows a
cross-correlation function corresponding to one of those orders. No
prominent peak is present in the cross-correlation. The middle panel
shows a simple average of the ten cross-correlation functions
($\rmn{SA}$). The correct peak is evidently present. It is
similarly present in the maximum-likelihood combined correlation --
$\rmn{ML}$, as the bottom panel shows. Since the cross-correlation
values are very small, $\rmn{DC}$\ and $\rmn{ML}$\ are virtually
identical. This simulation demonstrates the power of combining
cross-correlation functions in general. In this case no velocity could
have been measured for any individual order, but the cumulative effect
of the $10$ orders produced a detectable peak.

\begin{figure}
\resizebox{\hsize}{!}{\includegraphics{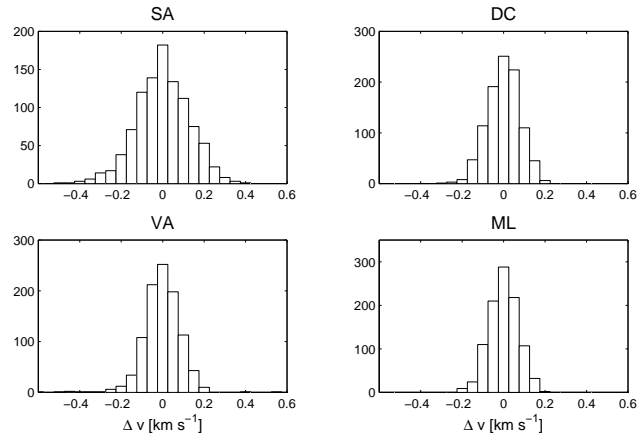}}
\caption{ 
The distribution of the peak location in each of the four combination
schemes.  See text for the details of the simulations. All four
histograms are drawn on the same horizontal scale for the sake of
comparison.}
\label{demo2}
\end{figure}

The next test shows the optimality of the maximum-likelihood scheme
($\rmn{ML}$) compared to the two averages $\rmn{SA}$\ and
$\rmn{DC}$. This is evident mainly in cases where the $S/N$ ratio is
strongly varying among the analysed spectra. To show that, noise was
added to the $10$ orders with standard deviations of $0.1 i$, where
$i$ is the order number. Figure \ref{demo2} shows the distribution of
the peak offset for each of the combination schemes, after repeating
the simulation for $1000$ times.  In addition, the Figure also
includes a histogram of the velocities derived by a weighted average
of the velocities obtained from each order separately
($\rmn{VA}$). The weighting was based on the individual error
estimates. Obviously, $\rmn{VA}$ can be calculated only if the correct
peaks can be identified in the cross-correlation functions of all
orders. The histogram corresponding to $\rmn{ML}$\ is evidently less
dispersed than all others.  Numerically, the four standard deviations
corresponding to $\rmn{SA}$, $\rmn{DC}$, $\rmn{VA}$ and $\rmn{ML}$
were $0.128$, $0.078$, $0.099$ and $0.068$ $\rmn{km\,s}^{-1}$.

An additional criterion to distinguish among combination schemes is
their ability to accentuate the correct peak relative to the spurious
peaks. This ability is important for detection of faint objects, with
low $S/N$. To test this ability, the typical height of the spurious
peaks was estimated by the standard deviation of the function in a
region that is safely distant from the correct peak.  Specifically,
this standard deviation was estimated on two segments of $500$ bins,
on both sides of the correct peak, at a distance of at least $100$
bins from it.  The mean ratios of the peak height to this standard
deviation were $36$, $58$ and $65$ for $\rmn{SA}$, $\rmn{DC}$\
and $\rmn{ML}$, correspondingly. Once again, we see that
$\rmn{ML}$\ provides the most prominent correct peak.

The above tests were repeated in numerous other conditions, in all of
them $\rmn{ML}$ was found to be the optimal scheme to combine
cross-correlations.

\begin{figure}
\resizebox{\hsize}{!}{\includegraphics{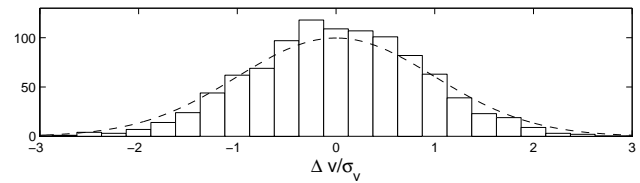}}
\caption{ 
Histogram of the ratio between the offset of the maximum-likelihood
combined correlations ($\Delta v$) and the estimated error
($\sigma_v$). The dashed line represent a zero-mean unit-variance
Gaussian distribution.}
\label{comberr}
\end{figure}

After establishing the optimality of the maximum-likelihood
combination scheme, its error estimate had to be validated, in a
similar fashion to the validation done for the single spectrum
case. Figure \ref{comberr} shows the $\Delta v/\sigma_v$ calculated
for the $\rmn{ML}$\ functions in the $1000$ simulations of $10$
orders. Once again, we see a very good agreement with a zero-mean
unit-variance Gaussian.

\section{Concluding remarks}
\label{concl}

This paper presents a novel approach for combining cross-correlation
functions. This approach is based on the relation of cross-correlation
to maximum-likelihood estimation.  Unlike previously suggested
approaches, this approach is not based on a linear combination of the
individual functions. Previous works looked for the optimal linear
combination, without considering other options
\citep[e.g.,][]{Con1985}. In cases where the scaling constants and the
signal-to-noise ratios are known, maximum-likelihood analysis will
indeed lead to a linear combination. However, exact estimate of the
signal-to-noise is difficult in real cases. The suggested combination
does not use any such external estimate.

Although the white gaussian noise model is assumed in most analyses of
the cross-correlation, it is obvious that reality is much more
complex. Thus, the noise level may depend on the intensity, like in
Poissonian noise, or vary across the spectrum because of illumination
effects. Additional complexities mainly include long-term trends and
spectral mismatch of the template to the analysed spectrum.  Attempts
to optimize radial-velocity measurement to deal with these problems
can be based also on maximum-likelihood theory.

The suggested approach was already applied successfully to the case of
HD\,41004, where $30$ {\small CORALIE} orders were combined in order to
detect a very faint companion of a K star \citep{Zucetal2003}. The
combined functions were not the conventional cross-correlation
functions, but two-dimensional correlation ({\small TODCOR}) functions
\citep{ZucMaz1994}. The successful detection of the faint companion
allowed a more detailed solution of the system and confirmed earlier
hypotheses of \citet{Sanetal2002}. This case demonstrates the power of
the technique to enhance the capabilities of traditional
cross-correlation and improve its ability to detect and measure very
faint objects, with very low $S/N$.

\section*{acknowledgements}

I am grateful to Tsevi Mazeh and Morris Podolak for useful discussions
and critical reading of the manuscript. I thank the Geneva planet
search group for the {\small CORALIE} spectrum of HD\,38858. This
research was supported by the Israeli Science Foundation (grant no.\
40/00) and the Jacob and Riva Damm Foundation.

\bibliographystyle{mn2e}
\bibliography{ref}

\end{document}